\documentclass[3p,times,twocolumn]{elsarticle}
 \biboptions{comma,sort&compress}
 
\usepackage{graphicx}
\usepackage{here}
\usepackage{ecrc}

\def\be{\begin{equation}}
\def\ee{\end{equation}}
\def\ba{\begin{eqnarray}}
\def\ea{\end{eqnarray}}
\def\bea{\begin{eqnarray}}
\def\eea{\end{eqnarray}}
\def\bes{\begin{subequations}}
\def\ees{\end{subequations}}
\newcommand{\A}{{\mathcal{A}}}
\volume{00}

\firstpage{1}

\journalname{Nuclear and Particle Physics Proceedings}

\runauth{}

\jid{nppp}

\jnltitlelogo{Nuclear and Particle Physics Proceedings}

\usepackage{amssymb}

\usepackage[figuresright]{rotating}

\begin{document}

\begin{frontmatter}

\title{QCD coupling which respects lattice restrictions at low energies
 $^*$}
 \cortext[cor0]{Talk given at 18th International Conference in Quantum Chromodynamics (QCD 17,  20th anniversary),  3 - 7 july 2017, Montpellier - FR}
 \author[label1]{C\'esar Ayala\fnref{fn1}}
   \fntext[fn1]{Speaker, Corresponding author.}
\ead{cesar.ayala@usm.cl}
\address[label1]{Department of Physics, Universidad T{\'e}cnica Federico Santa Mar{\'\i}a, Casilla 110-V, Valpara{\'\i}so, Chile\\ }

\pagestyle{myheadings}
\markright{ }
\begin{abstract}
We consider a phenomenologycal parametrization of the QCD running coupling which 
arises from the dispersion relation respecting the holomorphic properties of the physical 
QCD observables in the complex momentum plane. The parameters are fixed by the following requirements: 1) at enough high energies, it reproduces the underlying perturbative coupling, 2) at intermediate energy momenta, it reproduces the experimental semihadronic tau decay ratio, 
and 3) in the deep IR regime, it satisfies the qualitative properties coming from recent lattice results. 
Finally, we apply this new coupling to low-energy available experimental data. In particular, to Borel sum rules for $\tau$-decay, extracting the values of the dimension 4 and 6 condensates, to the V-channel Adler function, and to polarized Bjorken Sum Rule.
 
\end{abstract}

\end{frontmatter}
\section{The method: Constructing the Holomorphic Coupling}

We present a generalization/extension of the perturbative QCD running coupling 
under the assumption that it has a physical branch on the negative semiaxes of the 
$Q^2$-complex momenta plane, elsewhere it is a holomorphic function of $Q^2$. 
On the other hand, this coupling should satisfy the asymptotic freedom. These assumptions can be implemented via dispersion relation with the application of the Cauchy theorem to the integrand  $\A(Q'^2)/(Q'^2 - Q^2)$, i. e. 
\be
\A(Q^2) = \frac{1}{\pi} \int_{\sigma=M^2_{\rm thr}-\eta}^{\infty} \frac{d \sigma \rho_{\A}(\sigma)}{(\sigma + Q^2)} 
\qquad (\eta \to +0),
\label{Adisp}
\ee
where $\rho_{\A}(\sigma) \equiv {\rm Im} \mathcal{A}(-\sigma - i \varepsilon)$ is the discontinuity function (spectral function) of $\A$ along the cut. 

For different choices of $M_{\rm thr}^2>0$, we recover different known approaches. Between the most known are Fractional Analytic Perturbation Theory (FAPT) \cite{APT,BMS}; Massive Perturbation Theory (MPT) \cite{ShMPT} and $M\delta$ analytic QCD ($M\delta$anQCD) \cite{2danQCD,3danQCD1,3danQCD2}. In these  models, the threshold squared mass is 
\begin{equation}
M_{\rm thr}^2 = \left\{ \begin{array}{cl} 
-\Lambda_{\rm QCD}^2& ,{\rm pQCD}\\ 0& ,{\rm (F)APT}\\ 
m_{gl}^2-\Lambda_{\rm QCD}^2&, {\rm MPT} \\ \sim m_\pi^2&, M\delta {\rm anQCD}  \end{array}\right.
\end{equation}
Note that in pQCD and (F)APT $\rho_{\A}(\sigma) = \rho_{a}(\sigma)$, and MPT is defined from $\A_{\rm MPT}\equiv a(Q^2+m_{gl}^2)$. Here $a(Q^2)=\alpha_s(Q^2)/\pi$
In this work we will present and use a new coupling for the last case, i.e., 3$\delta$anQCD \cite{3danQCD1,3danQCD2}.

We define the unknown low-energy part of the integral (\ref{Adisp}) in the range $M_{\rm thr}^2<\sigma<M_0^2$ ($\sim1$ GeV$^2$) as $\Delta \A_{\rm IR}(Q^2)$. This integral has the same structure as a Stieltjes function \cite{Baker}. Then we can use a theorem that guarantees the convergence of a sequence of Pad\'es $[M-1/M]$ to $\Delta \A_{\rm IR}(Q^2)$ as $M\to\infty$. $[M-1/M]$ is a polynomial in $Q^2$ of power $M-1$ divided by a polynomial of power $M$. Therefore,

\bea
\Delta \A_{\rm IR}(Q^2) &\equiv&  \frac{1}{\pi} \int_{\sigma=M_{\rm thr}^2}^{M_0^2} \frac{d \sigma \rho_{\A}(\sigma)}{(\sigma + Q^2)} 
\label{M1M}
\nonumber\\
&=&  \sum_{j=1}^{M} \frac{{\cal F}_j}{Q^2 + M_j^2} \ .
\label{PFM1M}
\eea
And for $\sigma$ from $M_0^2$ to infinity, we recover the perturbative discontinuity $\rho_a(\sigma)$. That corresponds to 
\be
\rho_{\A}(\sigma) =  \pi \sum_{j=1}^{M} {\cal F}_j \; \delta(\sigma - M_j^2)  + \Theta(\sigma - M_0^2) \rho_a(\sigma) \ ,
\label{rhoA}
\ee
where $\Theta$ is the Heaviside step function. Then, the considered coupling $\A(Q^2)$ is parametrized as 
\be
\A(Q^2)=\sum_{j=1}^M \frac{{\cal F}_j}{(Q^2 + M_j^2)} + \frac{1}{\pi} \int_{M_0^2}^{\infty} d \sigma \frac{ \rho_a(\sigma) }{(Q^2 + \sigma)} \ .
\label{AQ2}
\ee
The coupling (\ref{AQ2}) has $2M+1$ free parameters ${\cal F}_j$, $M^2_j$ ($j=1,2,\ldots ,M$) and $M_0^2$.

In order to have a good estimation of the running coupling, the question is: how many delta functions are appropriate (sufficient) for reproduce the physics at $Q^2\lesssim 1$GeV$^2$?.

Before we answer this, let us show the main properties that a possible candidate for a new universal coupling should have:
\begin{enumerate}[(i)]
\item \label{HE} Reproduce the high-energy QCD phenomenology as obtained from perturbation theory. This requirement can be written as 
\be
\A(Q^2) - a(Q^2)   \sim \left( \frac{\Lambda^2_{\rm QCD}}{Q^2} \right)^{N_{\rm max}} \ ,
\label{Aadiff1}
\ee
for $|Q^2| > \Lambda^2_{\rm QCD}$, $N_{\rm max}>1$ sufficiently large, and 
\be
\A(M_Z^2)=\frac{\alpha_s(M_Z^2)}{\pi}=``{\rm world\ average}''
\label{AsymFree}
\ee
This is obtained from the world average value in the $\overline{\rm MS}$-scheme, e. g. 
$\alpha_s(M_Z^2,\overline{\rm MS})\approx0.1185$ \cite{PDG}. In Eq.~(\ref{AsymFree}) we should change the scheme according to our needs. 
\item \label{IE} Reproduce the intermediate-energy QCD phenomenology, by requiring that the main features of the semihadronic $\tau$-lepton decay physics be respected. Stated otherwise, we will require that the approach with the coupling $\A(Q^2)$ reproduce the experimentally suggested value of the V+A semihadronic $\tau$-decay ratio parameter $r^{(D=0)}_{\tau} \approx 0.20$ \cite{ALEPH2,DDHMZ}. This is the QCD part of the V+A $\tau$-decay ratio $R_{\tau} = \Gamma(\tau^- \to \nu_{\tau}{\rm hadrons}(\gamma))/\Gamma(\tau^- \to \nu_{\tau} e^- {\bar \nu}_e (\gamma))$, where the hadrons are strangeless ($\Delta S=0$) and the quark mass effects and other (small) higher-twist effects are subtracted, i.e., it is the dimension $D=0$ strangeless and massless part.
\item \label{LE} Satisfy some qualitative and/or quantitative properties of the coupling in the deep-IR region when $Q^2\to0$. In general, we have three different possibilities inspired by different physical/mathematical evidence. These are: IR-finite coupling (freezing); infinite effective coupling that reproduces confinement already at one loop level, and vanishing coupling inspired by lattice simulations. In this report, we will consider the last case, where the coupling should behave as $\A(Q^2)\sim Q^2$ at $Q^2\to0$.
\end{enumerate}

\section{Phenomenology: Fixing Parameters}

Now, we should take some decisions. The first (high-energy) condition (\ref{HE}) implies fixing the precision with respect to the underlying pQCD coupling, i.e., while $N_{\rm max}$ increases our precision increases too. We will use $N_{\rm max}=5$, which imply the following four equations (for the elimination of four free parameters). 
\be
\frac{1}{\pi} \int_{-\Lambda_{\rm QCD}^2}^{M_0^2} d \sigma \sigma^k \rho_a(\sigma)=\sum_{j=1}^3 {\cal F}_j M_j^{2 k}\ ,
\label{1u}
\ee
with $k=0,1,2,3$. The world average value will fix our $\Lambda_{\rm QCD}$ scale or equivalently, the underlying pQCD coupling $a(Q^2)$ and thus $\rho_a(\sigma)$.

The second (intermediate-energy) condition (\ref{IE}) will fix us one free parameter by the  semihadronic $\tau$-lepton decay physics.
The considered quantity $r^{(D=0)}_{\tau}$ is timelike, but it can be expressed theoretically, 
by using the Cauchy integral formula, by means of a spacelike quantity called (leading-twist and massless) Adler function $d(Q^2;D=0)$ \cite{Braaten,PichPra}:
\be
r^{(D=0)}_{\tau, {\rm th}} = \frac{1}{2 \pi} \int_{-\pi}^{+ \pi}
d \phi \ (1 + e^{i \phi})^3 (1 - e^{i \phi}) \
d(m_{\tau}^2 e^{i \phi};0) \ .
\label{rtaucont}
\ee
The Adler function $d(Q^2;D=0)$ is a derivative of the quark current correlator $\Pi$: $d(Q^2;D=0) = - 2 \pi^2 d \Pi(Q^2; D=0)/d \ln Q^2 - 1$, in the massless limit. Its perturbation expansion is known up to $\sim a^4$ \cite{d3} and rewritten in terms of the new coupling. The expansion in terms of the holomorphic coupling is different from the perturbative one due to nonperturbative nature of the theory, i. e., the analogs of the pQCD powers $a(Q^2)^n$ are specific functions $\A_n(Q^2)$ [$\not= \A(Q^2)^n$]
\bea
\lefteqn{d(Q^2;D=0)\equiv d(Q^2,\mu^2;D=0)_{\rm an}^{[4]} + {\cal O}(\A_5)}
\nonumber\\
& & = \A(Q^2)+d_1 \A_{2}(Q^2)+d_2 \A_{3}(Q^2)
\nonumber\\
&& + \ d_3 \A_{4}(Q^2) + {\cal O}(\A_5).
\label{dan}
\eea
The power analogs $\A_n(Q^2)$ from $\A(Q^2)$($=\A_1(Q^2)$) were constructed in general holomorphic theories from $\A(Q^2)$ by using renormalization group equations (RGE) Ref.~\cite{CV12} for integer $n$ and in Ref.~\cite{GCAK} for general real $n$.

The third (low-energy) condition (\ref{LE}) depends on what approach we will consider. We will take in this regime the information from the lattice simulations~\cite{LattcoupNf0} of the Landau gauge gluon $Z_{\rm gl}(Q^2)$ and ghost $Z_{\rm gh}(Q^2)$ dressing functions. These simulations were performed with large physical volume and high statistics, giving presumably reliable results in the low-momentum regime $0<Q^2 < 1 \ {\rm GeV}^2$. Then, we can obtain the lattice version of the coupling as 
\be
\A_{\rm latt.}(Q^2)  =  \A_{\rm latt.}(\Lambda^2)  \frac{Z_{\rm gl}^{(\Lambda)}(Q^2) Z_{\rm gh}^{(\Lambda)}(Q^2)^2}{{\widetilde Z}_1 ^{(\Lambda)}(Q^2)^2} \ ,
\label{Alatt}
\ee
where the value of the gluon-ghost-ghost vertex function is ${\widetilde Z}_1 ^{(\Lambda)}(Q^2)^2=1$ in the Landau gauge, and the UV cutoff squared $\Lambda$ is determined by the lattice spacing. The resulting lattice coupling (\ref{Alatt}) has two interesting features: it goes to zero as $\A_{\rm latt.}(Q^2) \sim Q^2$ when $Q^2 \to 0$ and has a a maximum at $Q_{\rm max}^2 \approx 0.135 \ {\rm GeV}^2$. These two properties will fix us two parameters.

Altogether we can adjust seven parameters of our coupling  (\ref{AQ2}), where four come from high energy, one from  intermediate and two from low-energy regime. This is equivalent to taking three delta functions in (\ref{rhoA}).

For practical implementation, we need the underlying pQCD coupling $a(Q^2)$, and thus $\rho_a(\sigma)$ in an explicit form if we want evaluate the integral in Eq.~(\ref{AQ2}). It is given by solving the $\beta$-function for a specific Pad\'e form \cite{GCIK} whose expansion gives the known MiniMOM coefficients $c_2({\rm MM},N_f=3)=9.2970$ and $c_3({\rm MM},N_f=3)=71.4538$ [the expansion of this Pad\'e $\beta$-function up to $\sim a(Q^2)^5$ reproduces the four-loop polynomial $\beta$-function]. This coupling involves Lambert function which can be easily implemented in Mathematica software. 
When comparing with lattice results, we must take into account the following relation between lattice MiniMOM (MM), \cite{MiniMOM} and $\overline{\rm MS}$-scheme scale convention
\bea
\frac{\Lambda_{\rm MM}}{\Lambda_{\overline{\rm MS}}} &=& 1.8968 \; ({\rm for} \; N_f=0); 
\nonumber\\
&=&1.8171 \; ({\rm for} \; N_f=3);
\label{LambdaMMMSbar}
\eea

\begin{table*}[hbt]
\setlength{\tabcolsep}{1.5pc}
\newlength{\digitwidth} \settowidth{\digitwidth}{\rm 0}
\catcode`?=\active \def?{\kern\digitwidth}
\caption{The seven parameters of the coupling $\A(Q^2)$: $M_j^2$ ($j=0,1,2,3$); ${\cal F}_j$ ($j=1,2,3$), both in [GeV$^2$]. These values are given for the representative case: $r^{(D=0)}_{\tau, {\rm th}}=0.201$ and $0.201 \pm 0.002$;  with $\alpha_s(M_Z^2;\overline{\rm MS}) = 0.1185 \pm 0.004$.}
\label{tab:effluents}
\begin{tabular*}{\textwidth}{llllllll}
\hline
$r^{(D=0)}_{\tau, {\rm th}}$ & $M_0^2$ & $M_1^2$  & $M_2^2$ & $M_3^2$ & ${\cal F}_1$ & ${\cal F}_2$  & ${\cal F}_3$\\
  \hline
$0.201$ & $8.719$ & $0.05$ & $0.247$ & $6.34$ & $-0.038$ & $0.158$  & $0.070$\\
$0.203$ & $9.254$ & $0.04$ & $0.329$ & $6.75$ & $-0.020$ & $0.143$  & $0.073$\\
$0.199$ & $8.211$ & $0.10$ & $0.143$ & $5.96$ & $-0.245$ & $0.361$  & $0.068$\\
  \hline
   \hline

\end{tabular*}
\end{table*}
\addtocounter{table}{-1}

In Table~\ref{tab:effluents} we present our results for the free parameters that fulfills the  three conditions (\ref{HE}), (\ref{IE}) and (\ref{LE}).

\begin{figure}[htb]
\vspace{9pt}
\centering\includegraphics[width=70mm]{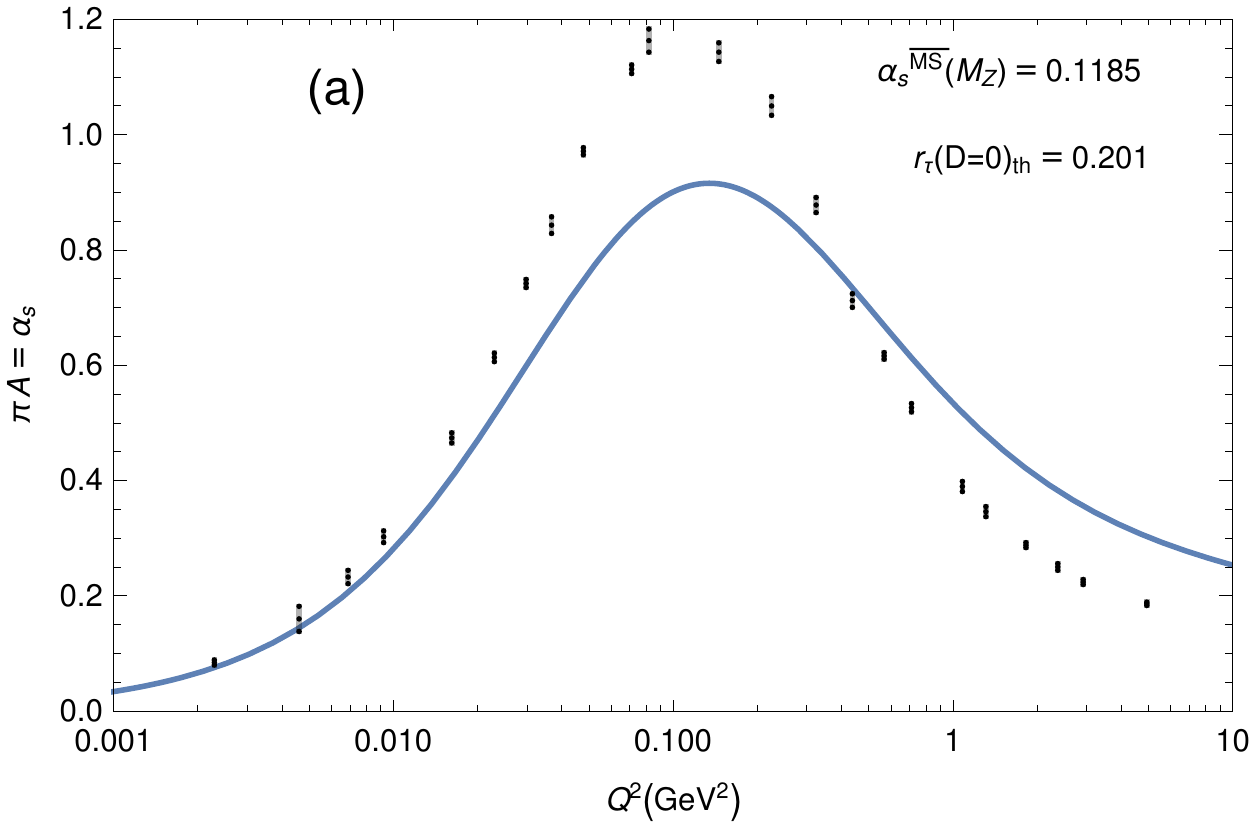}
\caption{The points represent the data obtained for the quenched lattice coupling from Ref.~\cite{LattcoupNf0} with their corresponding uncertainties. And the solid line our new coupling (\ref{AQ2}) with parameters given in Table \ref{tab:effluents}. We relate momenta in the MiniMOM (MM) lattice scheme to the usual ${\overline{\rm MS}}$-like scale \cite{3danQCD1,3danQCD2}.}
\label{fig:largenenough}
\end{figure}

In Fig.~\ref{fig:largenenough} we show the obtained ($N_f=3$) running coupling $\pi \A(Q^2)$ by solid the line, and the lattice $N_f=0$ calculations by points. In general our coupling agrees well with $\pi \A_{\rm latt.}(Q^2)$ at very low $Q^2$ ($Q \lesssim 0.01 \ {\rm GeV}^2$), and is lower than the lattice coupling near the maximum ($Q^2 \sim 0.1 \ {\rm GeV}^2$). We recall that we do not expect to have a good agreement between the theoretical and lattice coupling at $Q^2 \lesssim 0.1 \ {\rm GeV}^2$, but only a qualitative agreement. Even at higher $Q^2$ ($Q^2 > 1 \ {\rm GeV}^2$), there is a difference between $\pi \A(Q^2)$ and $\pi \A_{\rm latt.}(Q^2)$ of the higher-twist form $\sim\Lambda_{\rm QCD}^2$, and because we are working with $N_f=3$ while the lattice results \cite{LattcoupNf0} are for $N_f=0$. In fact, increasing $N_f$ in general decreases $\A_{\rm latt.}(Q^2)$, cf.~Fig.~5 of Ref.~\cite{LattcoupNf2}. Further, the lattice results concentrate on the deep IR regime, i.e., they had large lattice volume ($L \sim 10$ fm), but not small lattice spacing, which makes the lattice results \cite{LattcoupNf0,LattcoupNf2} unreliable at $Q^2\gtrsim1$GeV$^2$
 
\section{Applications}

In the present Section we will apply our coupling (\ref{AQ2}) with the corresponding parameters given in Table~\ref{tab:effluents} to some low-energy processes. 
Due to the condition (\ref{HE}), we can use OPE with $\A$-coupling in a way analogous to the OPE with pQCD $a$-coupling. In particular, due to Eq.(\ref{Aadiff1}) for $N_{\rm max}=5$, we can include in OPE with $\A$-coupling unambiguously the terms of dimensionality $D < 10$. The relevant programs in the implementation of this machinery are available and described online in Refs.~\cite{prgs,mathprg}.

\subsection{Borel Sum Rules to $\tau$-decay}

The application of dispersion relation to the polarization (current correlation) function $\Pi(Q^2)$ of the strangeless vector (V) and axial (A) currents gives us a holomorphic (analytic) function in the complex $Q^2$-plane, for $Q^2 \in \mathbb{C} \backslash (-\infty, -M_{\rm thr}^2]$ where the hadron production threshold mass is $M_{\rm thr}=M_1 \sim 0.1$ GeV. This quantity is then multiplied by any function $g(Q^2)$ [$\exp(Q^2/M^2)$ in the case of Borel Sum Rules] analytic in the entire complex $Q^2$-plane, 
and the Cauchy integral formula can be applied to the integral of $g(Q^2) \Pi_{V+A}(Q^2)$. With this, we arrive to the following relation
\be
{\rm Re} B_{\rm exp}(M^2) =  {\rm Re} B_{\rm th}(M^2) \ ,
\label{sr3o0}
\ee
where
\bea
\!\!\!B_{\rm exp}(M^2) &\equiv& \int_0^{\sigma_{\rm max}} 
\frac{d \sigma}{M^2} \; \exp( - \sigma/M^2) \omega_{\rm exp}(\sigma)_{V+A} \ ,
\label{sr3a}
\nonumber\\
\!B_{\rm th}(M^2) &\equiv&  \left( 1 - \exp(-\sigma_{\rm max}/M^2) \right)
+ B_{\rm th}(M^2;D\!=\!0)
\nonumber\\
&&+ 2 \pi^2 \sum_{n \geq 2}
 \frac{ \langle O_{2n} \rangle_{V+A}}{ (n-1)! \; (M^2)^n} \ ,
\label{sr3b}
\eea
and where the leading-twist contributions ($D=0$) is
\bea
\lefteqn{B_{\rm th}(M^2;D\!=\!0)=\frac{1}{2 \pi}\int_{-\pi}^{\pi}
d \phi \; d(\sigma_{\rm max} e^{i \phi};0) }
\nonumber\\
&&\times\left[ 
\exp \left( \frac{\sigma_{\rm max} e^{i \phi}}{M^2} \right) -
\exp \left( - \frac{\sigma_{\rm max}}{M^2} \right) \right] \ .
\label{BD0}
\eea
The total $D (\equiv 2 n) =2$ contribution in the OPE (\ref{sr3b}) is negligible, and we will include there the $D=4$ and $D=6$ terms. The advantage of the use of the Borel sum rules approach is that it is dominant in the low-$\sigma$ (IR) regime, and we can extract the gluon ($D=4$) and quark ($D=6$) condensates separately, depending on the choice of the complex argument [when $M^2=|M^2| \exp(i \pi/6)$, $D=6$ term in ${\rm Re} B_{\rm th}(M^2)$ is zero; and when  $M^2=|M^2| \exp(i \pi/4)$, the corresponding $D=4$ term is zero]. 

The experimental data used here are given by OPAL  \cite{OPAL} and ALEPH Collaborations \cite{ALEPH2,ALEPHfin}. Our combined fitting values of the condensates are \cite{3danQCD2}
\bea
\langle a GG \rangle&=&-0.0046\pm 0.0038\ [{\rm GeV}^4] ,
\label{aGGOPAL}
\\
\langle O_6 \rangle_{V+A}&=&0.00135\pm0.00039\ [{\rm GeV}^6].
\label{O6OPAL}
\eea

\begin{figure}[htb]
\vspace{9pt}
\centering\includegraphics[width=70mm]{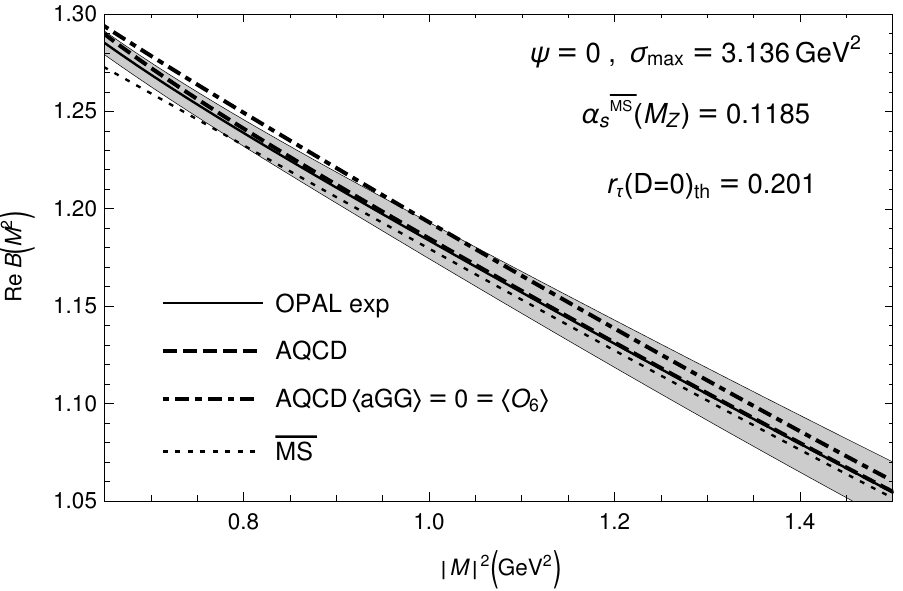}
\caption{Borel transforms ${\rm Re} B(M^2)$ for real $M^2 > 0$. The grey band represents the experimental results. For comparison, we show the fitted theoretical curve of $\overline{\rm MS}$ pQCD approach (dotted line). The theoretical curve given by our coupling ($\A$QCD) almost agrees with the central experimental (OPAL) curve.}
\label{FigPsi0}
\end{figure}

In Fig.~\ref{FigPsi0}, the curves for ${\rm Arg} M^2=0$ are presented, with the corresponding central values of the condensates obtained from OPAL Collaborations data (close to (\ref{aGGOPAL})-(\ref{O6OPAL}) values). We observe that our model applied in this case ($\A$QCD$+$OPE approach) agrees well with the (OPAL) experimental band in the entire presented $M^2$-interval, in contrast to the pQCD approach which agrees for the range $|M^2|\gtrsim 0.8$GeV$^2$.

\begin{figure}[htb]
\vspace{9pt}
\centering\includegraphics[width=70mm]{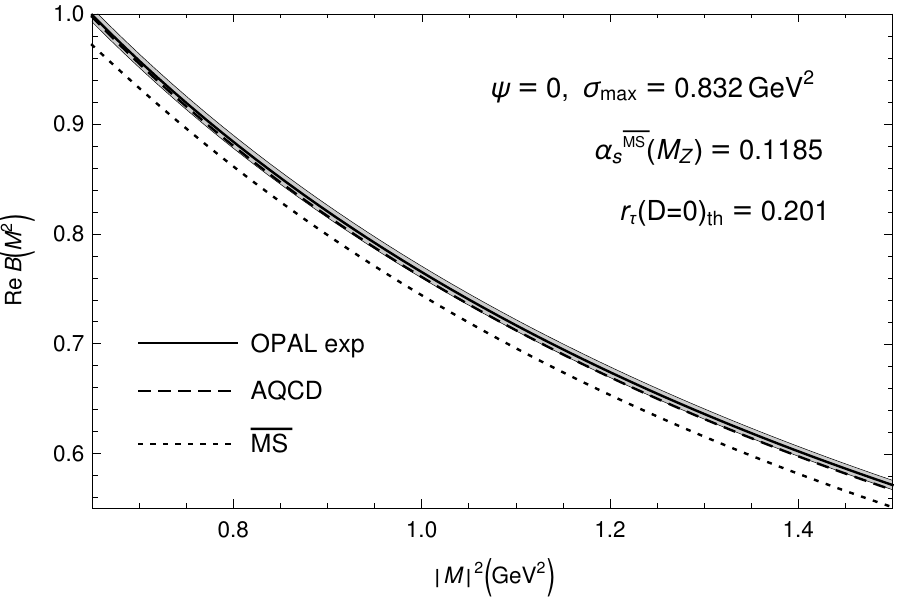}
\caption{As in Fig.~\ref{FigPsi0}, but now for a lower scale $\sigma_{\rm max}=0.832 \ {\rm GeV}^2$. The $\A$QCD curve (dashed) is inside the experimental band. Again, we show the $\overline{\rm MS}$ pQCD approach (dotted line) for comparison.}
\label{FigPsi0o26}
\end{figure}

We note that in the Borel sum rules we used $\sigma_{\rm max}=3.136 \ {\rm GeV}^2$ in the OPAL case, and $\sigma_{\rm max}=2.80 \ {\rm GeV}^2$ in the ALEPH case \cite{3danQCD1,3danQCD2}. We are interested in what happens when we decrease the value of $\sigma_{\rm max}$ while keeping the obtained original values of the condensates. In Fig.~\ref{FigPsi0o26} for instance, we show the case when $\sigma_{\rm max}=0.832 \ {\rm GeV}^2$; the $\A$QCD$+$OPE approach is significatively better than pQCD, the latter is located well outside the narrow experimental uncertainty band, and our approach remains inside in the whole presented range of $M^2$.
Similar results and conclusions given in this Section are obtained when using ALEPH Collaboration data \cite{3danQCD1,3danQCD2}.

\subsection{V-channel Adler function $\mathcal{D}_V(Q^2)$}

The V-channel Adler function ${\cal D}_V(Q^2)$ is related via dispersion relation with the production ratio $R(\sigma)$ for $e^+ e^- \to$ hadrons at the center-of-mass squared energy $\sigma$. The V-channel Adler function is
\bea
{\cal D}_V(Q^2) &\equiv&  - 4 \pi^2 \frac{d \Pi_V(Q^2)}{d \ln Q^2} 
 =  1 + d(Q^2;D=0) 
\nonumber\\
&&\qquad + 2 \pi^2 \sum_{n \geq 2}
 \frac{ n 2 \langle O_{2n} \rangle_V}{(Q^2)^n}  \ ,
\label{DV}
\eea
where $d(Q^2;D=0) $ is given by (\ref{dan}), and we estimate the values of the V-channel condensates from the values of the V+A channel condensates obtained in the previous Subsection. With these condensates, we can apply it using the relation $\langle O_4 \rangle_{V+A}=2 \langle O_4 \rangle_V$ ($=2 \langle O_4 \rangle_A$) for the $D=4$ condensates \cite{Braaten,PichPra}, and vacuum saturation approximation $\langle O_6 \rangle_{V+A}\approx-\frac{4}{7}\langle O_6 \rangle_V$ for the $D=6$ condensates \cite{3danQCD1,3danQCD2,Ioffe}.

\begin{figure}[htb]
\vspace{9pt}
\centering\includegraphics[width=70mm]{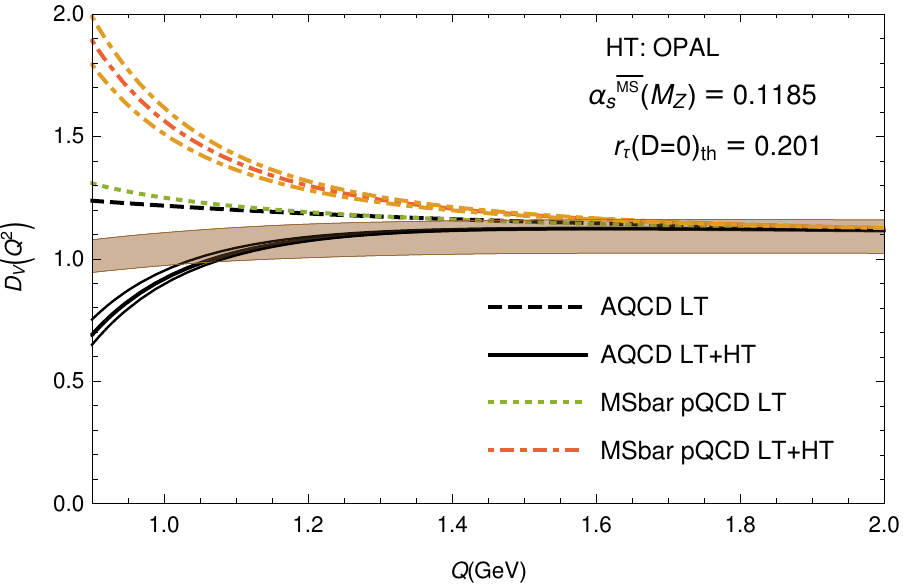}
\caption{The V-channel Adler function at $Q^2>0$. Experimental data are denoted by the grey band taken from \cite{NestBook}. The solid lines are our coupling (\ref{AQ2}) (or $\A$QCD), and the dash-dotted lines are in the $\overline{\rm MS}$ pQCD approach. The dashed line is the leading twist (LT) contribution in $\A$QCD, and the dotted line in $\overline{\rm MS}$ pQCD.}
\label{FigDVOPE}
\end{figure}

In Fig.~\ref{FigDVOPE}, the $\A$QCD+OPE approach gives results within the experimental band for all $Q^2$ down to  $Q^2 \approx 1 \ {\rm GeV}^2$, while the $\overline{\rm MS}$ pQCD+OPE only down to $Q^2 \approx 2.5 \ {\rm GeV}^2$. We stress that the incorporation of the lattice-motivated behavior for $\A(Q^2)$ at $|Q^2| \lesssim 0.1 \ {\rm GeV}^2$ influences significantly the behavior of $\A(Q^2)$ in the entire complex $Q^2$-plane, including in the regime of our principal interest, $|Q^2| \sim 1 \ {\rm GeV}^2$. The OPE series (\ref{DV}) is expected to fail always at $|Q^2| < 1 \ {\rm GeV}^2$

\subsection{Bjorken Sum Rule (BSR)} 

The polarized Bjorken sum rule (BSR) is defined as integral over the $x$-Bjorken of the nonsinglet combination of the proton and neutron polarized structure functions, i. e., 
\be
\Gamma_1^{p-n}(Q^2)=\int_0^1 dx \left[g_1^p(x,Q^2)-g_1^n(x,Q^2) \right]\ .
\label{BSRdef}
\ee 
BSR can be written in terms of a sum of two series, one coming from pQCD and the other from the higher-twist (HT) contributions dictated by the 
OPE \cite{BjorkenSR}
\be
\label{BSR}
\Gamma_1^{p-n}(Q^2)=\frac{g_A}{6}E_{\rm {NS}}(Q^2)+\sum_{i=2}^\infty
\frac{\mu_{2i}^{p-n}(Q^2)}{Q^{2i-2}}\ ,
\ee
 where the nucleon axial charge is $g_A=1.2723$ \cite{PDG2016}.We will include only the first HT term $\sim \mu_4^{p-n}$.

In our analysis it is convenient to exclude the elastic contribution, because the $Q^2$-dependence of the nonsinglet inelastic BSR in low-$Q^2$ regime is constrained by the Gerasimov-Drell-Hearn (GDH) sum rule \cite{GDHlow}, as was pointed out in \cite{PSTSK10}. 

The leading-twist (LT) contribution $E_{\rm {NS}}(Q^2)$ was calculated up to N$^3$LO contribution in \cite{nnnloBSR}.  

\begin{figure}[htb]
\vspace{9pt}
\centering\includegraphics[width=70mm]{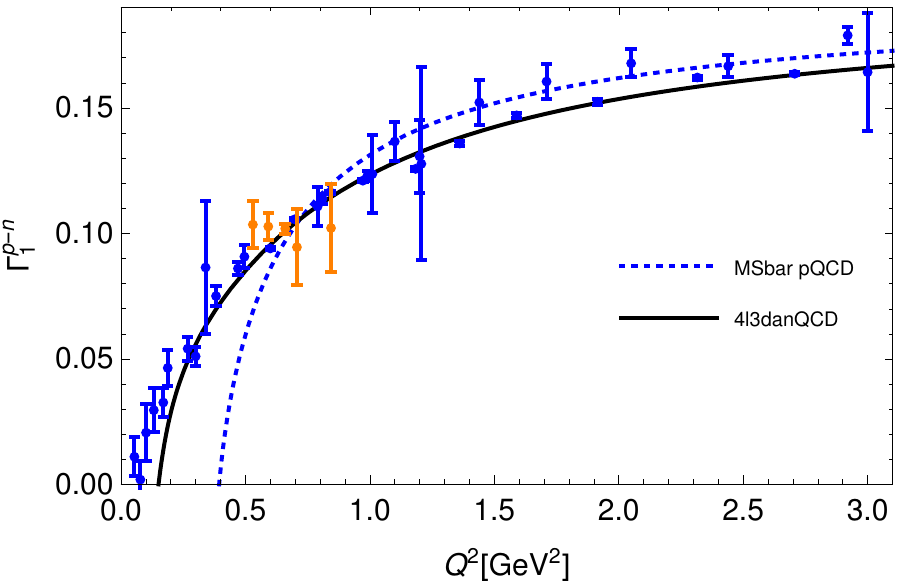}
\caption{Fits of JLAB and SLAC combined data \cite{dataBSR} on BSR $\Gamma_1^{p-n}(Q^2)$ as a function of $Q^2$, using (four-loop) $\overline{\rm MS}$ pQCD and our coupling $\A$ given by Eq.(\ref{AQ2}).}
\label{FigFitNf3}
\end{figure}

In analytic QCD approaches, the powers $a^{\nu}$ (where $\nu$ is not necessarily integer) get transformed to $\A_{\nu}$ (which is in general different from $\A^\nu$), according to the general formalism of Ref.~\cite{GCAK}. We apply it to the twist-4 term \cite{anBSR}
\be
\mu_{4,j}^{p-n}(Q^2)=\mu_{4,j}^{p-n}(Q_{\rm in}^2) \frac{\A_{\nu}^{(j)}(Q^2)}
{\A_{\nu}^{(j)}(Q_{\rm in}^2)}\ .
\label{HTQ2an}
\ee
where $\nu=1/8\beta_0$.
With the corresponding analytization of the HT term (\ref{HTQ2an}) and the implementation in the LT part, i.e., $a(Q^2)^n\mapsto\A_n(Q^2)$ in (\ref{BSR}), we can find a fit for $\mu_{4,j}^{p-n}(Q_{\rm in}^2)$ \cite{anBSR}. The resulting value at $Q_{\rm in}^2=1$GeV$^2$ is $\mu_{4,3\delta{\rm anQCD}}^{p-n}(1$GeV$^2)=-0.019$ and the corresponding plot is given in Fig.~\ref{FigFitNf3}.  We observe that the resulting fit describes almost the whole available experimental data (for $Q^2\gtrsim0.2$GeV$^2$) in contrast to pQCD which describes the data only for $Q^2\gtrsim0.7$GeV$^2$.

\section{Conclusions}

In this work we have presented a new QCD running coupling from their 
dispersive representation. Here we parametrize the IR regime of the spectral function with three delta functions. This allowed us to fulfill various physically motivated conditions, at high, intermediate and low momenta generating a holomorphic running coupling. The main feature is that in the deep-IR it behaves as $\A(Q^2)\sim Q^2$ as motivated by lattice calculations, and it reproduces the pQCD coupling at high-momenta.
Then we applied it to three different low-energy observables and we found that at $Q^2\sim1$GeV$^2$ scales our coupling is significantly better than pQCD+OPE approach. 

\section*{Acknowledgements}
This work was supported by FONDECYT (Chile) Postdoctoral Grant No.~3170116.

\end{document}